\documentstyle[sprocl]{article}

\begin{document}

\title{Electrons as quasi-bosons in magnetic white dwarfs}

\author{Jerzy Dryzek \footnote{e-mail: jerzy.dryzek@ifj.edu.pl}}

\address{Institute of Nuclear Physics, PL-31-342 Krak{\'o}w,
ul.Radzikowskiego 152, Poland}

\author{Akira Kato \footnote{e-mail: ak086@csufresno.edu},
Gerardo Mu{\~n}oz \footnote{e-mail: gerardom@csufresno.edu},
Douglas Singleton \footnote{e-mail: dougs@csufresno.edu}}

\address{Dept. of Physics, CSU Fresno, Fresno, CA 93740-8031, USA}

\maketitle

\abstract{A white dwarf star achieves its equilibrium from the balancing of
the gravitational compression against the Fermi degeneracy pressure of the
electron gas. In field theory there are examples ({\it e.g.} the
monopole-charge system) where a strong magnetic field can transform a boson
into a fermion or a fermion into a boson. In some
condensed matter systems ({\it e.g.} fractional quantum Hall systems) a
strong magnetic field can transform electrons into effective fermions, or
effective anyons. Based on these examples we investigate the possibility
that the strong magnetic fields of some white dwarfs may transform some
fraction of the electrons into effective bosons. This could have
consequences for the structure of highly magnetized white dwarfs. It would
alter the mass-radius relationship, and in certain instances one could
envision a scenario where a white dwarf below the Chandrasekhar limit could
nevertheless collapse into a neutron star due to a weakening of the electron
degeneracy pressure. In addition the transformation of electrons into
effective bosons could result in the electrons Bose condensing,
which could speed up the cooling rate of white dwarfs.}

\section{Introduction}

It has been known for a long time \cite{cha31,lan32} that certain
stars at the end of their life cycle reach an equilibrium where the
gravitational compression is balanced by the Fermi degeneracy pressure of
the electron gas. These white dwarf stars are theoretically
interesting since understanding their stability requires an
understanding of gravity, and of a quantum Fermi gas. The
stability argument for a white dwarf can be framed in terms of the Fermi
energy of the electrons versus their gravitational binding energy. The Fermi
energy for a relativistic electron is approximately 
\begin{equation}  \label{fe}
E_F \approx \frac{\hbar c N^{1/3}}{R}
\end{equation}
where $N$ is the number of electrons in the object, and $R$ is the radius of
the object. The gravitational energy per fermion is approximately 
\begin{equation}  \label{ge}
E_G \approx -\frac{GM m_n}{R}
\end{equation}
where $m_n \approx 1.67 \times 10 ^{-24}$ g is the nucleon mass,
and $M=N m_n$ is roughly the total mass of the star The total
energy is then $E_{tot} = E_F +E_G$. If the physical constants in
$E_F$ and $E_G$ are such that $E_{tot} >0$ then $E_{tot}$ can
be decreased by increasing $R$ and a stable situation is eventually reached
where the star is supported by its Fermi degeneracy pressure. If the
physical constants in $E_F$ and $E_G$ are such that $E_{tot} <0$ then
$E_{tot}$ decreases without bound by decreasing $R$ and no equilibrium
exists. The boundary between these two situations occurs when $\hbar c N
^{1/3} =G N m_n ^2$ which implies a maximum baryon number of $N_{max}
\approx (\hbar c / G m_n ^2) ^{3/2} \approx 2 \times 10 ^{57}$ and a
maximum total mass of $M_{max} \approx N_{max} m_n \approx 1.5
M_{\bigodot}$. This simple argument gives an approximation of the
Chandrasekhar mass limit for white dwarfs. Crucial to this argument (or more
rigorous versions) is that the electrons should behave as fermions in order
to give rise to the Fermi degeneracy pressure. 
In lower dimensional field theories there are examples, such
as the sine-Gordon model in one space and one time dimension, 
where bosonic and fermionic degrees of freedom can be taken as
dual or interchangeable \cite{col75,man75}. These lower
dimensional examples can be extended to $3+1$ dimensions \cite{lut79}.
In Abelian and non-Abelian field theories \cite{jac76,has76}
there are configurations where, through the action of a
magnetic field, the statistics of the system
can be transformed ({\it i.e.} the system can be fermionic even
though all the fields involved in its construction are bosonic).
There are also certain condensed matter systems,
where fermions can be converted into effective fermions
or effective bosons. The fractional quantum Hall effect
\cite{pra90} offers one such example, where electrons in 2D systems in the
presence of a large magnetic flux can act as effective fermions \cite{jai89}
or effective bosons \cite{gir87,dun89} depending on which
picture/approach one uses. Here we look at the possibility that in highly
magnetized white dwarfs a similar transformation may occur for some fraction
of the electrons of the star. Converting some
fraction of the electrons of the star into quasi-bosons would mean that they
would no longer be involved in giving rise to the Fermi degeneracy pressure:
the $N$ in Eq. (\ref{fe}) would be reduced by the fraction of
the electrons which are converted to effective bosons. This weakening of
the Fermi degeneracy pressure could result in a white dwarf with a mass
below the Chandrasekhar limit collapsing into a neutron star. Further, this
collapse of an otherwise stable white dwarf to a neutron star, might occur
without a supernova. This could offer an explanation for certain pulsar
systems which have planets orbiting them \cite{wol92}. If the neutron star
formed via a supernova then the original planets of the progenitor
star should have been blown out of the system. Also transforming electrons
into quasi-bosons could result in these electrons Bose condensing, which
could speed up the cooling rate of the white dwarf.

\section{Electron field angular momentum in a magnetic white dwarf}

In this section we will give a simple picture for how an electron inside a
highly magnetized white dwarf can be transformed in an effective boson.
We will frame our arguments for this transformation in terms of the
field angular momentum of the system rather than the additional
phase factor that arises in the exchange of the charge/magnetic flux
composites. Various 2D \cite{wil82} and 3D \cite{gol76}
examples have found that both approaches lead to an equivalent
understanding of the change in statistics of these composite systems.

The idea that a fermion can be converted to a boson, or a boson into a
fermion via a magnetic field, can be found in certain non-Abelian field
theories. These examples involve placing a charged particle in the
vicinity of one of the finite energy magnetic
monopole solutions \cite{tho74,pol74} which
exist in these field theories. This composite of particle
plus monopole possesses a field angular momentum which gives the system
statistics the opposite of what is naively expected \cite{gol76}. In the
original example of Refs. \cite{jac76,has76} both the monopole and
charged particle were bosons, while the bound state composite was a
fermion. A related phenomenon in 2D was discussed by \cite{wil82} where the
electric charge combined with a magnetic flux tube quantum to give a
composite which had different statistics from naive expectations. For the 2D
case one can have objects ({\it anyons}) which have statistics that fall
anywhere between fermions and bosons \cite{gir90}. In three spatial
dimensions, however, one can show that on topological grounds \cite{gir90}
only bosons or fermions occur. These arguments leave open the
possibility of boson to fermion, or fermion to boson transformations
in three spatial dimensions (it is the latter case that we are
interested in). Recently this transformation of
fermions into composite bosons, or bosons into composite fermions
has been experimentally observed in the effectively 2D fractional
quantum Hall systems \cite{tsu82}. Since magnetic white dwarfs are 3D
objects the fermion to quasi-boson transformation that we are proposing
here has more in common with the 3D monopole/charge systems.

Based on these field theory and condensed matter examples we propose that
electrons in a highly magnetic white dwarf could be transformed into
quasi-bosons via the star's magnetic field. Specifically the magnetic field
could combine with the electric field of the electron to give rise to a
field angular momentum. If this field angular momentum is of the correct
magnitude ({\it i.e.} some half-integer multiple of $\hbar$) then the
combination of the electrons' spin plus the field angular momentum will
result in an effective boson which will not contribute to the Fermi
degeneracy pressure. This would affect the equilibrium of the white dwarf.
It would change its mass-radius relationship, so that for a given mass one
would have a smaller radius than for a white dwarf with a smaller magnetic
field. It might even be possible for such a magnetic white
dwarf to collapse into a neutron star despite being
below the Chandrasekhar limit.

The angular momentum carried in the electric and magnetic fields can be
written as \cite{jac75} 
\begin{equation}  \label{1}
\mathbf{L}_{em} = {\frac{1 }{4 \pi c}} \int \mathbf{r \times (E \times B)}
d^3 \mathbf{r}
\end{equation}
Now we consider an electron located at the origin in a uniform magnetic
field whose direction is taken to define the z-axis. We will assume that
near the location of the electron the magnetic field $\mathbf{B} = B_0 {\hat
{\mathbf{z}}}$, where $B_0$ is the magnitude of the magnetic field. In a
white dwarf the electrons form a conducting gas with a background lattice of
the positive nuclei. The electric field of the electron will be screened
past a certain distance, $R_{sc}$. This screening effect is taken into
account by using a Yukawa potential for the electron's electric field
\begin{equation}  \label{poten}
\phi (\mathbf{r}) = \frac{e \exp (-r/R_{sc})}{r}
\end{equation}
The potential in Eq. (\ref{poten}) gives an E-field of 
\begin{equation}  \label{efield}
\mathbf{E} = -\frac{e \exp (-r/R_{sc})}{r^2} \left[ 1 + \frac{r}{R_{sc}} 
\right] \mathbf{\hat r}
\end{equation}
Carrying out the calculation of the field angular momentum which results
from this combination of magnetic and electric fields yields
\begin{equation}  \label{fam}
\mathbf{L}^{em} = {\frac{2 e B_0 R_{sc}^2 }{c}}\mathbf{\hat z}
\end{equation}
The direction of field angular momentum is in the $\mathbf{\hat z}$
direction, which is determined by the direction of the external magnetic
field. Despite the singularity in the Coulomb field of the electron, the net
field angular momentum does not diverge. The $B_0 R_{sc} ^2$ part of this
expression is proportional to the magnetic flux, $\Phi$ ``trapped'' by the
electron. As will be discussed below $R_{sc}$ for a typical white
dwarf is in general small -- on the order of $10^{-10}$ cm. Both the
spin of the electron and the field angular momentum will be localized
within a sphere of radius $10^{-10}$ cm. Thus it is physically
reasonable to view this screened electron plus trapped magnetic
flux as a composite system whose total angular momentum is a combination
of the electron's spin plus the field angular momentum -- $L_{tot} =
Spin + Field$. (One can contrast this with the monopole/charge system
where the field angular momentum {\em density} is not necessarily
well localized around the charge and yet is still taken as being part of
the total angular momentum of the composite system).
If $L^{em}_z$ in Eq. (\ref{fam}) takes on some half-integer
multiple of $\hbar /2$ then the total angular momentum should be that of a
boson (such arguments were originally used in \cite{sah36,wil49}
to arrive at the Dirac quantization condition
for electric charge). As in the monopole/charge system
of Refs. \cite{jac76,has76},
the electron will behave as an effective boson and not contribute to the
degeneracy pressure. A similar picture ({\it i.e.} the combining of a
field angular momentum with the spin of the electron) was proposed 
\cite{dry00} to give an alternative, complementary picture of the
fractional quantum Hall effect where the electrons are transformed into
effective fermions by the external field. Here, in contrast, we want the
magnetic field magnitude, $B_0$, and/or screening distance, $R_{sc}$, to
take on values which yield $L^{em} _z = (n + \frac{1}{2}) \hbar$
so that $L^{tot} _z = L^{spin} _z + L^{em} _z = m \hbar$
($n,m$ are integers) thus making effective bosons. Highly
magnetized white dwarfs can have surface magnetic field strengths up to
$10 ^8$ or $10 ^9$ gauss. We are interested in the magnetic
field strengths in the interior for which unfortunately there is no direct
experimental determination. However, as an estimate we will take the
interior field to be two orders of magnitude greater than the surface fields
({\it i.e.} interior fields of $10 ^{10}$ or $10 ^{11}$ G).
It is not uncommon to postulate interior magnetic fields as large as
$10^{13}$ G \cite{suh00}).

The screening length of the electric field inside a white dwarf can be
estimated as \cite{sha83} 
\begin{equation}  \label{screen}
R_{sc} =\sqrt{\frac{E_F}{6 \pi e^2 Z n_0}}
\end{equation}
where $E_F$ is the Fermi energy, $n_0$ the number density of the ions that
form the lattice, and $Z$ is the number of positive charges of each ion. For
these quantities we take the following values which are typical for carbon
white dwarfs -- $Z=6$, $E_F = 0.6 \times 10^{-6}$ ergs, $n_0 =1.7 \times
10^{29}$ cm$^{-3}$ (this corresponds to an electron number density of $1.0
\times 10^{30}$ cm$^{-3}$) -- which leads to $R_{sc} = 3.7 \times 10^{-10}$
cm. Given this screening length and requiring that $L^{em}$ from Eq. (\ref
{fam}) equal $\hbar/2$ yields $B_0 = 3.6 \times 10^{11}$ G.
This magnetic field strength is possible in the interiors
of highly magnetized white dwarfs \cite{suh00}. Thus some fraction of the
electrons within a highly magnetized white dwarf could be transformed into
quasi-bosons in analogy to what occurs in magnetic charge/electric charge
composites, or to the quasi-bosons in certain condensed
matter systems with strong magnetic fields.

The question that is not addressed in the above analysis is what fraction of
the electrons are transformed into quasi-bosons? In the present scenario
the field angular momentum needs to take on odd-integer multiples of $\hbar
/2$ in order to turn the electrons into quasi-bosons. The field angular
momentum depends on both the background magnetic field, $B_0 \mathbf{\hat z}$,
and the screening distance, $R_{sc}$. If one takes these two quantities as
varying with position inside the white dwarf, then as one moves around the
interior of the white dwarf some regions will have the correct values to $B_0
$ and $R_{sc}$ in order for the transformation to occur. As an illustrative
example consider a magnetic white dwarf where it is assumed that $R_{sc}$ is
constant throughout the star and takes on the value given after Eq. (\ref
{screen}) -- $R_{sc} \approx 3.7 \times 10^{-10}$ cm. Take the magnetic
field at the surface to be on the order of $10^9$ G and
let its magnitude increase
linearly with radius, from the surface of the star, to an interior, central
value of $10^{13}$ G. Taking the radius of the white dwarf to be $5 \times
10^8$ cm then implies that at a radius of $4.8 \times 10^8$ cm ({\it i.e.}
just below the surface) the magnetic field will take on the correct value in
order to generate a field angular momentum of $\hbar /2$. All the electrons
in the spherical shell at this radius will be transformed into quasi-bosons.
As one continues further into the star a new radius will be reached where
the magnetic field strength will increase to the point that the field
angular momentum will become $3 \hbar /2$ again transforming the electrons
into quasi-bosons. Going all the way into the center of the white dwarf one
encounters a series of spherical shells where the electrons are changed into
effective bosons. The result being that the electron degeneracy pressure is
reduced, altering the mass-radius relationship for the star.

Since the magnetic fields discussed here are large relative to those that
can be found in a laboratory, one should also ask if quantum modification of
the magnetic field strength alter any of the above analysis. If the quantum
corrections reduced the magnetic field strength this would tend to
increase the point at which the transformation of electrons into
quasi-bosons took place. However, for magnetic field strengths on the
order of $\simeq 10^{15}$ G one can show  \cite{mun96} that the
quantum corrections are on the order of $\simeq 1 \%$. For the
magnetic field strengths considered here the quantum corrections
can be safely ignored.

\section{Physical consequences of quasi-bosonic electrons}

There are two main physical consequences which could result from the
transformation of electrons into bosons. First, as mentioned in
the previous sections, the mass-radius relationship would be altered in such
a way that the radius would be smaller than what would normally be expected.
This decreasing of the radius, due to the transformation of the electrons
into effective bosons, is in the opposite direction of another effect of a
strong magnetic field. In Ref. \cite{suh00} it is
shown that for magnetic white dwarfs the pressure of the electron gas is
increased through a strong magnetic field, tending to make the radius for a
given mass larger than without the magnetic field. However, this effect only
starts to become important for magnetic field strengths in the range of 
$10^{12} - 10^{13}$ G. The magnetic field strengths that transform electrons
into effective bosons are one or two orders of magnitude less than this, so
that the decrease in radius due to the transformation of the electrons into
bosons should begin before the
increase in radius due to the mechanism of Ref.
\cite{suh00}. One interesting possible consequence of this weakening of
the electron degeneracy pressure through the fermion $\rightarrow$ boson
mechanism, is that it may allow magnetic white dwarfs which are slightly
below the Chandrasekhar limit to collapse, without a supernova,
into a neutron star. Consider a magnetic white dwarf which is just below the
Chandrasekhar limit, and should therefore be stable against further
collapse. Take the interior magnetic field of the white dwarf to be in the
range of $10^{12} - 10^{13}$ G so that the fermion $\rightarrow$ bosons
transformation could occur in the manner described at the end of the last
section ({\it i.e.} the magnetic field magnitude varies linearly from its
surface value to its interior value so that there will be a series of
concentric shells where the electrons become quasi-bosons). The Fermi
degeneracy pressure could be reduced enough so that the star's inward
gravitational pressure would be slightly larger, allowing the white dwarf
to slowly collapse, without a supernova explosion, into a neutron star.
This collapse scenario may offer an explanation of the planet-pulsar
systems \cite{wol92}. In these systems one has up to three planets
orbiting a pulsar. If this pulsar formed in a supernova collapse then
the initial planets should have been blown out of the system, which
is taken to imply that these planets must have formed after the supernova.
However, if the original star collapsed slowly due to the
fermion $\rightarrow$ boson transformation, then the
current planets might be the original planets of the star.

Second, the cooling period of these magnetic white dwarfs could be
accelerated, since the quasi-bosonic electrons could Bose condense. In Ref. 
\cite{nag00} the idea was advanced that certain white dwarfs might be Bose
condensed systems, with the positively charged ions being the bosons which
are condensing. For an ideal Bose gas the condensation temperature is 
\begin{equation}  \label{boset}
T_0 = \frac{2 \pi \hbar ^2}{m k} \left( \frac{n}{2.612} \right) ^{2/3}
\end{equation}
$k$ is Boltzmann's constant and $n$ is the number density. When an ideal
Bose gas drops below this critical temperature it will Bose condense.
Assuming that this is a first order transition \cite{nag00} implies that an
energy of 
\begin{equation}  \label{deltae}
\Delta E = L M
\end{equation}
is released by the condensation. $L$ is the latent heat of
condensation and $M$ is the total mass of the condensing system. This
increases the cooling rate of the star. In Ref. \cite{nag00} the condensing
particles are the positively charged ions of the white dwarf. This
resulted in a condensation temperature from Eq. (\ref{boset}) in the range
$10^4 - 10^6$ K. Using the fermion to boson transformation proposed above one
can think to apply this Bose condensation idea to electrons
which sit in an appropriately strong magnetic field. Because of the inverse
relationship between $T_0$ and the mass of the condensing particle in Eq.
(\ref{boset}), and since the electron is three to four orders of magnitude
lighter than the ions, the critical temperature of these quasi-bosonic
electrons to Bose condense is in the range of $10^7 - 10^9$ K. The Bose
condensation of these electrons would occur at much higher temperatures as
compared to the condensation of the ions. One might think that the energy
released by the electron condensation would be smaller since in this case $M$
from Eq. (\ref{deltae}) would be the mass of all the electrons which were
transformed into quasi-bosons. However, the smaller mass of the electrons
relative to the ions is compensated for by the fact that the latent heat of
condensation is proportional to $T_0$ : $L \propto T_0$ (see
Ref. \cite{nag00}). Since
the $T_0$ for the quasi-bosonic electrons is increased by the same factor
that $M$ is decreased, these two effects cancel in the expression for
$\Delta E$. One final difference between a Bose condensation of ions versus
a Bose condensation of quasi-bosonic electrons, is that for the electrons
only those which are transformed will Bose condense, while for the ions
they all condense. However the number of electrons is larger than the number
of ions ({\it e.g.} for a carbon white dwarf there are 6 electrons for
every ion), thus even if only some fraction of the electrons are transformed
they could nevertheless give a similar contribution to $\Delta E$ as in the
ion condensation scheme of Ref. \cite{nag00}.
For the carbon white dwarf example
if $15 \%$ of the electrons are transformed into bosons then the number of
quasi-bosonic electrons will be only a little less than the number of ions.
Thus the Bose condensation mechanism of
Ref. \cite{nag00} could be applied to
the transformed electrons, but the critical temperature at which Bose
condensation occurs would be much higher, because of the smaller electron
mass. This would have the effect of accelerating the cooling time for such
highly magnetic white dwarfs.

\section{Discussion}

In this paper we have proposed a speculative mechanism whereby some fraction
of the electrons within a highly magnetized white dwarf can be transformed
by a sufficiently strong magnetic field into effective bosons. The
system of an electron placed in a uniform magnetic field has a field angular
momentum, whose magnitude is proportional to the strength of the magnetic
field and the square of the screening distance of the electric field of the
electron. For certain values of the magnetic field and screening distance
this field angular momentum can take on half-integer values,
which combined with the half-integer spin of the electron leads
to the combined object of electron plus trapped magnetic
flux having an integer angular momentum. It is postulated
that this transforms the electron into a quasi-boson. Similar
transformations of fermions into bosons or bosons into fermions via the
action of a magnetic field can be found in the charge/monopole systems
studied in particle physics \cite{jac76,has76} and in
the fractional quantum Hall systems in condensed matter physics
\cite{jai89,gir87,dry00}.
There are two physical consequences that could result from turning some of
the electrons inside a highly magnetized white dwarf into effective bosons :
First, the Fermi degeneracy pressure would be decreased, altering the
mass-radius relationship of the white dwarf so that for a given mass the
radius would be smaller than if the magnetic field were absent.
Such highly magnetic white dwarfs, with quasi-bosonic electrons,
might be able to collapse into neutron stars despite being below
the Chandrasekhar limit. The second possible physical
consequence is a faster cooling rate of such magnetic white dwarfs,
via Bose condensation. This is similar to the idea of
Ref. \cite{nag00} where an
accelerated cooling rate was proposed through the Bose condensation of the
integer spin nuclei of the white dwarf. The present work differs from this
in that the objects which are taken as Bose condensing are the transformed
quasi-bosonic electrons.

It would appear that the above mechanism is not applicable to pure,
ideal neutron stars. Even though the magnetic field strength of a neutron
star can be several orders of magnitude larger than that of a white dwarf, a
neutron carries no charge, and therefore no field angular momentum would be
generated by placing the neutron in a magnetic field. This of course assumes
the approximation of treating the neutron as a fundamental, chargeless
object. As one goes to smaller distance/larger energy scales ({\it i.e.}
as one considers neutron stars with increasing densities) there may be a
transition where one needs to describe the matter, not in terms of neutrons,
but in terms of a gas of charged quarks. At this point one might again
consider applying the mechanism discussed in this paper, but now the
magnetic field would be transforming the fermionic quarks into quasi-bosons.

\section*{Acknowledgment}

We wish to thank Dr. Fred Ringwald for comments and discussions.

\end{document}